\documentstyle[12pt]{article}
\begin{document}

\baselineskip =15.5pt
\pagestyle{plain}
\setcounter{page}{1}

\begin{titlepage}

\begin{flushright}
\end{flushright}
\vfil
\begin{center}
{\huge Addenda and corrections to work done on the path-integral approach
to classical mechanics}

\end{center}

\vfil
\begin{center}
{\large E.Gozzi and M.Regini}\\
\vspace {1mm}
Dipartimento di Fisica Teorica, Universit\`a di Trieste, \\
Strada Costiera 11, P.O.Box 586, Trieste, Italy \\ and INFN, Sezione 
di Trieste.\\
\vspace{3mm}
\end{center}

\vfil

\begin{center}
{\large Abstract}
\end{center}

\noindent
In this paper we continue the study of the path-integral formulation
of classical mechanics and in particular we better clarify, with respect 
to previous papers, the geometrical meaning of the variables 
entering this formulation.
With respect to the first paper with the same title, we  {\it correct}
here the set of transformations for the auxiliary variables $\lambda_{a}$.
We prove that under this new set of transformations the Hamiltonian 
~${\widetilde{\cal H}}$, appearing in our path-integral,
is an exact scalar and the same for the Lagrangian.
Despite this different transformation, the variables $\lambda_{a}$~maintain the
same operatorial meaning as before but on a different functional space.
Cleared up this point we then show that the space spanned  by the whole set of 
variables~($\phi, c, \lambda,\bar c$) of  our path-integral is  
the cotangent bundle to the {\it reversed-parity} tangent bundle of the
phase space~${\cal M}$ of our system and it is indicated as $T^{\star}(\Pi
T{\cal M})$.
In case the reader feel uneasy with this
strange {\it Grassmannian} double bundle, we  show in this paper 
that it is possible to build a different path-integral made only of 
{\it bosonic}  variables. These turn out to be the coordinates of
~$T^{\star}(T^{\star}{\cal M})$ which is the double cotangent bundle of 
phase-space.
\vfil
\end{titlepage}
\newpage
\newcommand{\be}{\begin{equation}}
\newcommand{\ee}{\end{equation}}
\newcommand{\bea}{\begin{eqnarray}}
\newcommand{\eea}{\end{eqnarray}}
\def \HT{{\widetilde{\mathcal H}}}
\def \LT{{\widetilde{\mathcal L}}}
\def \bi{\bibitem}
\def \ci{\scite}
\newcommand{\quattrova}{($\phi^{a},c^{a},\lambda_{a},{\bar c}_{a}$)}
\newcommand{\treva}{($\phi^{a},c^{a},{\bar c}_{a}$)}
\newcommand{\scite}{~\cite}

\section{Introduction}
In this paper we continue the study of the path-integral formulation
of Hamiltonian classical mechanics started in ref.\scite{En1} and 
continued in ref.\scite{En2}. Here  in particular we provide
a better  understanding of the geometrical meaning of the space spanned
by the variables~($\phi^{a},c^{a},\lambda_{a},{\bar c}_{a}$) entering
that path-integral\scite{En1}. Instead of reviewing here all the previous work,
we suggest that the reader not familiar 
with the formalism read  refs.\scite{En1,En2} before embarking in this paper.

In refs.\scite{En1,En2} we showed that,
somehow magically, our path-integral was giving us tools which
provided the full Cartan calculus\scite{Mrs} on symplectic spaces but
we failed in giving  a geometrically clean understanding of the full space 
we were working on. In those papers there were some incorrect statements
regarding the nature of the full space.
It took us some time to fully understand which space those
variables,($\phi,c,\lambda,{\bar c}$), were spanning  because both $\lambda_{a}$ 
and ${\bar c}_{a}$
transformed as basis of the vector fields so both of them
seemed to span the tangent bundle $T{\cal M}$ to the phase-space ${\cal M}$
whose coordinates we called $\phi$. Later on things started
clearing up . 

The first things that got clear was a puzzle we had on the
role of the variables $\lambda_{a}$. In ref.\scite{En2} we had supposed
that the $\lambda_{a}$~ transform, under a (symplectic)
diffeomorphism of the phase-space variables ~$\phi^{a}$, as the basis
of the vector fields.  The transformations  are
indicated in eqs. (A9) of ref.\scite{En2}. But at the same time we 
knew that the~$\lambda_{a}$~, under the time evolution,
change as indicated in eq.(3.16) of ref.\scite{En1}. Now time-evolution
is a particular symplectic diffeomorphism so (3.16) of ref.\scite{En1}
should fall in the class of transformations (A9) of ref.\scite{En2}.
But this is not the case, as it is easy to see inspecting the two
transformations. In (3.16) of ref.\scite{En1} there is in fact an extra piece
containing the anticommuting variables $c, \bar c$ which is not present
in (A9) of ref.\scite{En2}. Which  is the correct transformation? The 
extra piece present in (3.16) seems to ruin the clean geometrical
character of the $\lambda_{a}$~ which,we thought,  were the basis of the vector
fields. We had that  belief because in ref.~\cite{En1} we proved (see eq.
(3.32)) that at the operatorial level~
$\lambda_{a}=-i{\partial\over\partial\phi^a}$. Because
of that belief we postulated the transformations (A9) of ref.\scite{En2}.

We prove here in section 2 that the correct transformation (under
any diffeomorphism) is the one of ref.\scite{En1} with the extra piece made
of Grassmannian variables\footnote{We will see that it is easy to
generalize the transformation from the time evolution to any
general diffeomorphism.}.This extra piece does not spoil
the geometrical meaning of ~$\lambda$ as the  basis of the vector
fields. The reason being that one should consider which is the correct
space of functions on which the vector fields act as a derivative.
A further indication that
those of ref.\scite{En1}  should be the correct transformations comes from the 
fact that under those transformations the Hamiltonian~$\widetilde{\mathcal H}$
~entering our path-integral is an exact scalar while it was not so
 under the other transformation
as shown in Appendix A of ref.\scite{En2}.

Another issue which we have clarified in this paper, and on which we made
wrong  statements  in ref.\scite{En1,En2}, is the manner to build 
{\it symmetric} tensors.
If we use the partial geometrical results already obtained
in ref.\scite{En1}, we can show that any  form-valued {\it anti-symmetric}
tensors
can be built out of the variables ($\phi,c,\bar c$). Motivated by this result
we then thought that the {\it symmetric} tensors could be built
using the  $\lambda$ in place of the ${\bar c}$. But it is not so unless 
we provide a symplectic connection. Details are provided in  
section 3. Several calculational details are confined in an appendix
at the end of the paper.

Cleared up this fact we turned in section 4 to study the geometrical
nature of our space. The  thing we  notice is  that the $c^{a}$, together
with the $\phi^{a}$, make not really a cotangent bundle $T^{\star}{\cal M}$
to phase-space\footnote{We erroneously stated that in refs.\scite{En1,En2}}
but a space  known in the literature
as {\it reversed parity} tangent bundle\scite{Scw} which is indicated
as~$\Pi T{\cal M}$.
Putting all this  together 
we  show in section 3 that the space  spanned by the full set of
variables ($\phi,c,\lambda,\bar {c}$) is the 
cotangent bundle to the {\it reversed-parity} tangent bundle of  phase-space
indicated as $T^{\star}(\Pi T{\cal M})$.  

The reader may feel a little up uneasy 
with the strange {\it Grassmannian } double bundle we produced with our 
variables. To lift this uneasiness we provide in section 5 of this paper 
a purely  bosonic path-integral whose 8n-variables are the coordinates 
not of a Grassmannian bundle but of a more natural
$T^{\star}(T^{\star}{\cal M})$~ which is the double cotangent bundle to
phase-space.

\section {Correct Transformations of $\lambda$}

As we said in the introduction the main goal of this paper is to clarify
the geometrical meaning of the space spanned by the variables \quattrova.
The first puzzle we encountered was the following: both $\lambda_{a}$ and
${\bar c}_{a}$ transformed under a (symplectic) diffeomorphism of $\phi$
(which are the coordinates of the phase-space ${\cal M}$) as the basis of
vector fields\scite{En1,En2}. For $\lambda$~ this was clear from the
fact that in the operatorial formulation associated to our path-integral
one had $\lambda_{a}=-i{\partial\over\partial \phi^{a}}$, while for
${\bar c}$ we transformed it that way because\scite{En1}~${\bar c}={\partial
\over\partial c^{a}}$ and $c^{a}$ transforms as a form. We thought
that it was a little bit strange that we had two copies (one Grassmannian
and one commuting) of the basis of vector fields.
The first idea was that  maybe we had done a mistake in considering 
$\lambda$ as the basis of vector field. This thought come from
considering how the $\lambda_{a}$~ change under the simplest of (symplectic)
diffeomorphisms that is the Hamiltonian evolution. The Lagrangian associated
to our path-integral~\cite{En1} is:
\be
\label{eq:lagran}
\LT=\lambda_{a}\left[{\dot \phi}^{a}-\omega^{ab}\partial_{b}H\right]+
i{\bar c}_{a}\left[\delta^{a}_{b}\partial_{t}-\omega^{ac}
\partial_{c}\partial_{b}
H\right]c^{b}
\ee
where $H$ is the Hamiltonian\footnote{Repeated indices are meant as summed
all through this paper} of our system in the phase-space~${\cal M}$.
From the $\LT$~above we can derive the equation of motion for $\lambda$
\footnote{Note that the term with the
$c$~and ${\bar c}$ variables
has a different sign than in \scite{En1}. The correct one is
the one written here in this paper.}:
\be
\label{eq:evol-lambda}
\partial_{t}\lambda_{a}=-\lambda_{i}\omega^{ij}\partial_{j}\partial_{a}H-i
{\bar c}_{i}\omega^{ij}\partial_{j}\partial_{b}\partial_{a}H c^{b}
\ee
From (\ref{eq:evol-lambda}) one can derive the transformation under an infinitesimal time 
interval~$\Delta t$: 
\be
\label{eq:infi-lambda}
\lambda^{\prime}_{a}=\left[\delta^{b}_{a}-\omega^{bj}\partial_{j}
\partial_{a}H\Delta t\right]\lambda_{b}-i{\bar c}_{i}
\omega^{ij}\left(\partial_{j}\partial_{b}\partial_{a}H\Delta t\right)c^{b}
\ee
Let us compare this transformation with the one of eq.(A9) of ref.\scite{En2}
which is the correct one if $\lambda$~were the basis of the vector fields
and it is:
\be
\label{eq:lambda-errata}
\lambda^{\prime}_{a}={\partial\phi^{b}\over\partial\phi^{\prime a}}\lambda_{b}=
\left[\delta^{b}_{a}+\partial_{a}(\omega^{bc}\partial_{c}G)\right]\lambda_{b}
\ee
where $G$~is related to the infinitesimal (symplectic) transformation
$\phi^{\prime a}=\phi^{a}-\varepsilon^{a}(\phi)$~ as follows: 
$\varepsilon^{e}=\omega^{ef}(\partial_{f} G)$. 
For the infinitesimal
time evolution of eq.(\ref{eq:infi-lambda}), the $G$ would be $G=-H\Delta t$.
In that case eq.(\ref{eq:lambda-errata}) would become 
\be
\label{eq:primo-pezzo}
\lambda^{\prime}_{a}=\left[\delta^{b}_{a}-\omega^{bj}\partial_{j}
\partial_{a}H\Delta t\right]\lambda_{b}
\ee
We see immediately that this is not the correct
time-evolution given by eq.(\ref{eq:infi-lambda}) the second piece
containing ${\bar c}$ and $c$ is totally missing.

The extra piece, under a generic (symplectic) diffeomorphism generated
by the $G$ of eq.(\ref{eq:lambda-errata}), would turn this 
equation  into:

\be
\label{eq:lambda-giusto}
\lambda^{\prime}_{a}=\lambda_{a}+\omega^{ij}(\partial_{j}\partial_{a}G)
\lambda_{i}+i{\bar c}_{i}\omega^{ij}(\partial_{j}\partial_{b}\partial_{a}G)c^{b}
\ee

Let us first see the effect of this   extra piece on the transformation
properties of $\LT$ of eq.(\ref{eq:lagran}) and on the associated 
$\HT$ which is

\be
\label{eq:hami}
\HT=\lambda_{a}\omega^{ab}\partial_{b}H+i{\bar c}_{a}\partial_{b}(\omega^{ac}
\partial_{c}H)c^{b}
\ee

Using eq.(\ref{eq:lambda-errata}) the variation\footnote{Of course we
change not only $\lambda$ but also the other variables. These ones
are changed as in eq. (A9) of ref.\scite{En2}.} of $\HT$  was given in
ref.\scite{En2} (under eq.(A9)) and 
was\footnote{With ${\widetilde{\cal H}}^{\prime}(\phi^{\prime})$
we indicate an ${\widetilde{\cal H}}$ of the same "form" (in the new variables)
as the one of eq.(\ref{eq:hami})
but with $H$ replaced by the canonically transformed $H^{\prime}$ indicated
in eq.(A9) of ref.\scite{En2}.}:

\be
\label{eq:variaz-errata}
\Delta{\HT}={\widetilde{\mathcal H}}^{\prime}(\phi^{\prime})-\HT(\phi)
=ic^{b}{\bar c}_{a}\omega^{fc}(\partial_{c}H)\partial_{b}\partial_{f}
(\omega^{ae}\partial_{e}G)+O(G^{2})
\ee 

With the new transformation of eq.(\ref{eq:lambda-giusto}) there will
be an extra piece in the variation of $\HT$. This extra piece
will come only from the part of $\HT$ containing $\lambda$.It is easy to find 
this extra variation\footnote{Of course we keep transforming the
other variables $\phi, c,{\bar c}$ as in eq. (A9) of ref.\scite{En2}}:

\be
\label{eq:variaz-extra-HT}
\Delta_{extra}\HT=(\Delta_{extra}\lambda_{a})\omega^{ab}\partial_{b}H=i{\bar c}
_{i}\omega^{ij}(\partial_{j}\partial_{b}\partial_{a}G)c^{b}\omega^{al}
\partial_{l}H
\ee

Let us now sum the two variations (\ref{eq:variaz-errata}) and
(\ref{eq:variaz-extra-HT})

\begin{eqnarray}
\Delta_{tot}{\HT} & = & \Delta{\HT}+\Delta_{extra}{\HT} \\
& = & ic^{b}{\bar c}_{a}\omega^{fc}(\partial_{c}H)\omega^{ae}(\partial_{e}
\partial_{b}\partial_{f}G)+ i{\bar c}_{i}\omega^{ij}(\partial_{j}\partial
_{b}\partial_{a}G)c^{b}\omega^{al}\partial_{l}H \nonumber 
\end{eqnarray}
If, in the first piece on the RHS of this equation, we relabel the indeces
in the following manner $a\rightarrow i$,$f\rightarrow a$,$c\rightarrow l$,
$e\rightarrow j$, and in the second piece we bring the last $c$-variable
in front, we get:

\be
\label{eq:finale}
\Delta_{tot}{\HT}=ic^{b}{\bar c}_{i}\omega^{al}(\partial_{l}H)\omega^{ij}
(\partial_{j}\partial_{b}\partial_{a}G)-ic^{b}{\bar
c}_{i}\omega^{al}\partial_{l}
H\omega^{ij}(\partial_{j}\partial_{b}\partial_{a}G)=0
\ee
of course this is true up to terms $O(G^{2})$ but this does not matter
because  G is infinitesimal in an infinitesimal transformation.

So we have proved that the transformations (\ref{eq:lambda-giusto})
leave the $\HT$ invariant or better ${\widetilde{\cal H}}$~behaves as a scalar under a 
(symplectic) diffeomorphism. 
To complete the calculations we will now show that the Lagrangian $\LT$
is actually invariant\footnote{We keep using the word "{\it invariant}"
but it is wrong, we mean "{\it scalar}". "{\it Invariant}" would in-fact
mean $\LT^{\prime}(\phi,\cdots)=\LT(\phi, \cdots)$~ while "{\it scalar}"
means $\LT^{\prime}(\phi^{\prime},\cdots^{(\prime)})=\LT(\phi, \cdots)$.
"{\it Invariant}" for a Lagrangian would imply there is a conserved quantity
under those transformations and, of course,  it is not the case here ! We are
only checking the {\it covariance} property of our Lagrangian.}
without the need to use the equations of motion\footnote{That was a need we had
in ref.\scite{En2} eq.(A10).}.
In appendix A of ref.\scite{En2} we proved that under the transformation
(\ref{eq:lambda-errata}) the $\LT$ changes as

\be
\label{eq:variaz-lagra}
\LT^{\prime}=\LT+ic^{b}{\bar c}_{a}{\partial^{2}(\omega^{ac}\partial_{c}G)\over
\partial\phi^{f}\partial\phi^{b}}{\dot \phi}^{f}-ic^{b}{\bar c}_{a}
{\partial^{2}(\omega^{ac}\partial_{c}G)\over \partial\phi^{f}\partial\phi^{b}}
\omega^{fd}\partial_{d}H
\ee

The third piece on the RHS of the equation above comes from the term on the
RHS of eq.(\ref{eq:variaz-errata}).
It is the same in fact but with the opposite sign because $\LT$ is related to
$-\HT$ under the Legendre transformation. As we know that
under the new transformation (\ref{eq:lambda-giusto}) the $\HT$
will not have that extra piece, the same will happen for $\LT$. We have now
only to check if even the second piece on the RHS of eq.(\ref{eq:variaz-lagra})
will disappear under the new transformation (\ref{eq:lambda-giusto}).
Actually in $\LT^{\prime}$ we have the kinetic piece 
$\lambda_{a}^{\prime}{\dot \phi}^{\prime a}$ containing $\lambda$
and  not present in $\HT^{\prime}$. This will produce, thanks to the new 
transformation (\ref{eq:lambda-giusto}), an extra term which may cancel 
the second piece in the RHS of~(\ref{eq:variaz-lagra}). This is in fact so. 
The extra piece
produced is: $i{\bar c}_{i}\omega^{ij}(\partial_{j}\partial_{b}
\partial_{a}G)c^{b}{\dot\phi}^{a}$ which is exactly the opposite of the second
piece on the RHS of (\ref{eq:variaz-lagra}). So this proves that, with the
new transformation (\ref{eq:lambda-giusto}) we obtain, instead of
(\ref{eq:variaz-lagra}), the following:

\be
\label{eq:variaz-giusta-lagra}
\LT^{\prime}(\phi^{\prime},\cdots)=\LT(\phi, \cdots)
\ee

So it seems that the correct transformation for $\lambda$ is
(\ref{eq:lambda-giusto}) which has the extra-piece containing the
Grassmannian variables $c,{\bar c}$.

The reader may  suspect that the extra piece present in the 
transformations of $\lambda$ is needed only if we insist in having a new
${\widetilde{\cal H}}^{\prime}$ with  
the same form of the ${\widetilde{\cal H}}$ of eq.(\ref{eq:hami}), that means if
we insist in having the 
Lie-derivative of an {\it Hamiltonian} flow\scite{En1,En2}.
We will prove that this is not the case.
If we start from the generic 
Lie-derivative $l_{V}$ of an arbitrary vector-field $V$ 
(not an Hamiltonian one)
it can be written\scite{Mrs} in a coordinate free way  as
$l_{V}\equiv di_{V}+i_{V}d$~where $d$ is the exterior derivative and
$i_{V}$~is the interior contraction with the vector field $V$.
As $l_{V}$ is a coordinate free expression it must be  {\it invariant} under
any change of coordinates (or diffeomorphism). That Lie-derivative
in our notation\scite{En1} can be written as:

\be
\label{eq:lie-deriv}
{\widetilde{\cal H}}_{V}\equiv\lambda_{a}V^{a}+i{\bar c}_{a}
\partial_{b}V^{a}c^{b}
\ee

and\footnote{We see that, if $V^{a}=\omega^{ab}\partial_{b}H$, we end up
in our Lie-derivative of the Hamiltonian flow (\ref{eq:hami}). Of course not
any $V$ can be written that way, only the Hamiltonian ones.} we  will show that in order to have "invariance" 
of this generic 
Lie-derivative under any  diffeomorphism we need the extra piece in the 
transformation of $\lambda$.

Let us transform the variables entering entering ${\widetilde{\cal H}}_{V}$ 
as:

\bea
\label{eq:trans}
{\phi^{\prime}}^{a} & = & \phi^{a}-\varepsilon^{a}(\phi)\\
\label{eq:trans1}
{c^{\prime a}} & = & c^{a}-c^{b}\partial_{b}\varepsilon^{a}\\
\label{eq:trans2}
{\bar c}^{\prime}_{a} & = & {\bar c}_{a}+
{\bar c}_{b}\partial_{a}\varepsilon^{b}\\
\label{eq:trans3}
\lambda_{a}^{\prime} & = & \lambda_{a}+\lambda_{b}\partial_{a}\varepsilon^{b}\\
\label{eq:trans4}
{V^{\prime}}^{a} & = & V^{a}-V^{b}\partial_{b}\varepsilon^{a}
\eea

where $\varepsilon^{a}(\phi)$ is a vector field generating
a generic diffeomorphism\footnote{It is easy to see that if
$\varepsilon^{a}=\omega^{ab}\partial_{b}G$ we would end up with our
transformations (A9) of ref.\scite{En2}. These last transformations are called
symplectic
diffeomorphisms
because they would lead to a new $V^{\prime}$~ which could  still be written
as an Hamiltonian vector field 
${V^{\prime}}^{a}=\omega^{ab}\partial_{b}H^{\prime}$ with only a different $H$.
With the calculations shown in the first part of this chapter we can say that
(if we use the new transformation for $\lambda$)~the Lie-derivative
of an Hamiltonian vector field is invariant under symplectic diffeomorphism.
What we will prove next is that we need that different transformation of
$\lambda$ also to prove that a generic Lie-derivative is coordinate
free (or invariant) under a generic diffeomorphism.}

It is a simple calculation to show that under the transformations
above  ${\widetilde{\cal H}}_{V}$ is not invariant:

\be
\label{eq:noninv}
\Delta {\widetilde{\cal H}}_{V}\equiv{\widetilde{\cal H}}_{V^{\prime}}^{\prime}-
{\widetilde{\cal H}}_{V}=-i{\bar c}_{a}V^{m}\partial_{b}\partial_{m}\varepsilon
^{a}c^{b}
\ee
If in eq.(\ref{eq:trans3}) we transform $\lambda$ with the extra ghost
piece as:

\be
\label{eq:uffa}
\lambda_{a}^{\prime}=\lambda_{a}+\lambda_{b}\partial_{a}\varepsilon^{b}+i
{\bar c}_{i}(\partial_{a}\partial_{b}\varepsilon^{i})c^{b}
\ee

then we get that $\Delta{\HT}_{V}=0$. So even for a generic Lie-derivative under a generic diffeomorphism
we have to change $\lambda$ in a different way in order to have
invariance of the Lie-derivative. This proves that this change
of $\lambda$ is a fundamental property and it is not
related to having treated only the Lie-derivative of  Hamiltonian flows
under symplectic diffeomorphisms.

We have anyhow now to reconcile what we proved above 
with the fact that, at the operatorial level, we had got\scite{En1}
\be
\label{eq:lambda-operatoriale}
\lambda_{a}=-i{\partial\over\partial \phi^{a}}
\ee

We say that we have to reconcile these two facts because we thought that
eq.(\ref{eq:lambda-operatoriale})~would imply that the $\lambda_{a}$, 
being proportional to ${\partial\over\partial\phi^{a}}$, would transform as the basis of
vector fields i.e. as  indicated in eq.(\ref{eq:lambda-errata}) and not
as in eq.(\ref{eq:lambda-giusto}).
 
The way to reconcile this is the following : ${\partial\over\partial\phi^{a}}$
would transform as a vector field if it were applied to functions
only of $\phi$, i.e. $F(\phi)$, but it would transform differently
if it were applied to functions of $\phi$~and $c$, i.e. $F(\phi,c)$.
We will now show that, if applied to these last functions, they would
transform as the $\lambda_{a}$~ do in eq.(\ref{eq:lambda-giusto}).

As explained in appendix A of ref.\scite{En2}, under  the same  (symplectic)
diffeomorphism which is applied to $\lambda$,~the $\phi$ and $c$ 
transform as:
\bea
\label{eq:diffeo}
\phi^{\prime a} & = & \phi^{a}-\omega^{ab}\partial_{b}G \\
\label{eq:diffeo1}
c^{\prime a} & = & \left[\delta^{a}_{b}-\omega^{ac}\partial_{c}
\partial_{b}G\right]c^{b}
\eea
Note that the transformed $c^{\prime}$ depend on $\phi$ via $G$.
Let us now take a function of $\phi$ and $c$, i.e. $F(\phi,c)$ and let
us transform its arguments.

\be
\label{eq:transf-function}
F(\phi,c)=F\left[\phi(\phi^{\prime}),c(\phi^{\prime},c^{\prime})\right]\equiv
{\cal S}(\phi^{\prime},c^{\prime})
\ee

If we now apply ${\partial\over\partial\phi^{\prime}}$ on ${\cal S}$,
we have

\bea
\label{eq:deriv-func}
{\partial\over\partial\phi^{\prime}}{\cal S}(\phi^{\prime}, c^{\prime}) & = &
{\partial\over\partial\phi^{\prime}} F\left[\phi(\phi^{\prime}), c(\phi^{\prime}
,c^{\prime})\right]=\\
& = & {\partial
F\over\partial\phi}{\partial\phi\over\partial\phi^{\prime}}+
{\partial F\over\partial c}{\partial c\over\partial\phi^{\prime}}
\eea
Comparing the two RHS of the above equations we can say that:

\be
\label{eq:deriv-sole}
{\partial\over\partial\phi^{\prime a}}={\partial\phi^{b}\over\partial
\phi^{\prime a}}{\partial\over\partial\phi^{b}}+{\partial c^{b}\over
\partial\phi^{\prime a}}{\partial\over\partial c^{b}}
\ee

Using the operatorial correspondence described in ref.\scite{En1}:
${\partial\over\partial \phi^{\prime a}}=i\lambda_{a}^{\prime}$,~\break
${\partial\over\partial\phi^{a}}=i\lambda_{a}$~and~ ${\partial\over\partial
{c^{b}}}={\bar c}_{b}$, we can re-write eq.(\ref{eq:deriv-sole})
as

\be
\label{eq:coord-free}
i\lambda_{a}^{\prime}=i{\partial\phi^{b}\over\partial\phi^{\prime
a}}\lambda_{b}+{\partial c^{b}\over\partial\phi^{\prime a}}{\bar c}_{b}
\ee

One could say that in general ${\partial c^{b}\over\partial\phi^{\prime a}}=0$
because $c$ and $\phi^{\prime}$ are independent coordinates.
Actually it is not so because as, we saw in eq.(\ref{eq:diffeo1}), the
transformed $c$ depend on $\phi$  and viceversa the original $c$ depends
on the transformed $\phi$. So ${\partial c^{b}\over\partial
\phi^{\prime a}}\neq 0$.  

Let us now proceed to see if eq.(\ref{eq:coord-free}) is the same as
eq.(\ref{eq:lambda-giusto}). From eqs.(\ref{eq:diffeo})~and~(\ref{eq:diffeo1}) 
we can derive their inverse which are:

\bea
\label{eq:invers}
\phi^{a} & = & \phi^{\prime a}+\omega^{ab}\partial_{b}G(\phi^{\prime})\\
c^{a} & = &
\left[\delta^{a}_{b}+\omega^{ac}\partial_{c}\partial_{b}G\right]c^{\prime b}
\eea

and from here we get:

\bea
\label{eq:deriv-inv}
{\partial\phi^{b}\over
\partial \phi^{\prime a}} & = & \delta^{b}_{a}+\omega^{bi}\partial_{i}
\partial_{a}G \\
{\partial c^{b}\over \partial \phi^{\prime a}} & = & \omega^{bc}\partial_{c}
\partial_{i}\partial_{a}G c^{\prime i}
\eea

Inserting these expressions into eq.(\ref{eq:coord-free}) we obtain

\be
\lambda^{\prime}_{a}=\left[\delta^{b}_{a}+\omega^{bi}\partial_{i}\partial_{a}
G\right]\lambda_{b}-i(\omega^{bc}\partial_{c}\partial_{i}\partial_{a}G)c^{i}
{\bar c}_{b}
\ee

where we have replaced $c^{\prime}$ with $c$ because  we keep only terms which 
are first order in the infinitesimal~$G$. Bringing now ${\bar c}$ in front on the second 
term of the RHS of the equation above we get:

\be
\lambda^{\prime}_{a}=\left[\delta^{b}_{a}+\omega^{bi}\partial_{i}\partial_{a}G
\right]\lambda_{b}+i{\bar c}_{b}\omega^{bc}\partial_{c}\partial_{i}\partial_{a}
G c^{i}
\ee

which is exactly the transformation for $\lambda$ we had in eq.
(\ref{eq:lambda-giusto}). So this proves that operatorially $\lambda_{a}$ act as ${\partial\over
\partial\phi^{a}}$ but over the functions $F(\phi,c)$ and this in turn
implies that the {\it base-space} we should consider first is the one made of
$(\phi,c)$.

\section{Symmetric tensors}
\par
Having cleared the issue of the transformations of the $\lambda$
let us now turn to another topic which was only
briefly mentioned in refs.\scite{En1,En2}. The topic is the following:
a generic function of the coordinates $(\phi,c,\bar c)$, i.e. 

\be
\label{eq:form-val-tens}
{\cal F}={\cal F}^{ab\cdots
m}_{i,j,\cdots n}(\phi)c^{i}c^{j}\cdots c^{n}{\bar c}_{a}{\bar c}_{b}\cdots {\bar
c}_{m}
\ee
represent a {\it form-valued anti-symmetric tensor field}.
This is so because\scite{En1} the $c^{a}$ are forms while the ${\bar c}_{a}$
transform as vectors~(\ref{eq:trans2}). So the strings of ${\bar c}_{a}$ in
(\ref{eq:form-val-tens}) gives
to the ${\cal F}$ the character of an anti-symmetric multivector \scite {Vienna}
while the strings of $c$ makes it a multi-form. Stripping it of the
$c$ variables we get a simple anti-symmetric multivector field:
\be
\label{eq:tenso}
{\cal T}={\cal T}^{ab\cdots m}(\phi){\bar c}_{a}{\bar c}_{b}\cdots {\bar c}_{m}
\ee
Somehow this statement was already present
in refs.\scite{En1,En2} where forms and tensors were written with our 
variables $c^{a}$, ${\bar c}_{a}$ and all the Cartan calculus 
was given in details.  
\par
In those same references\scite{En1,En2} we gave hints that {\it symmetric}
tensors could be built by replacing in (\ref{eq:tenso}) the ${\bar c}_{a}$
(which are anticommuting and provide the anti-symmetric character of 
the tensor ${\cal T}$) with the variables $\lambda_{a}$ which, being commuting,
appeared as the natural objects to build symmetric tensors. We {\it wrongly} 
thought that objects like:
\be
\label{eq:tensosim}
{\cal S}={\cal S}^{ab\cdots m}(\phi){\lambda}_{a}\lambda_{b}\cdots\lambda_{m}
\ee
would transform as symmetric tensors. The reason why this is wrong
is because the $\lambda_{a}$ do not transform as indicated in
eq.(\ref{eq:lambda-errata}), i.e, as vectors, which would make the ${\cal S}$
a {\it symmetric} multivector, but they transform as indicated in eq.
(\ref{eq:lambda-giusto}). So, as the $\lambda_{a}$ are not anymore vectors,
the ${\cal S}$ are not anymore symmetric multivectors. The same would happen
if we use the operatorial representation of $\lambda$~i.e.
 $\lambda_{a}=-i{\partial\over\partial \phi^{a}}$. For example, 
let us build a symmetric object with two indices  which would be:
${\cal S}^{ab}{\partial\over\partial\phi^{a}}
{\partial\over\partial\phi^{b}}$
It is well known that, while a single derivative 
${\partial\over\partial\phi^{a}}$
would transform as a vector component, a double derivative
${\partial\over\partial\phi^{a}}{\partial\over\partial\phi^{b}}$ would not
transform as the product of two vector components, unless we introduce
a connection\scite{novi}. 
\par
So the problem is to check if in our formalism we can find a manner
to build not only {\it anti-symmetric}  but also  {\it symmetric}
tensors  and eventually even {\it mixed} ones. For the moment let us focus
our attention on the symmetric ones.
\par
We see that without a connection there is no manner to make symmetric even a
2-tensor. So the new crucial ingredient to pull in our construction
seems to be a connection $\Gamma$.
A connection is an object needed to do the parallel transport of vectors
\scite{novi}. Under a diffeomorphism $\phi^{\prime a}=
\phi^{\prime a}(\phi)$ of the space ${\cal M}$, the connection does not 
transform  as  a tensor but in the following manner:
\be
\label{eq:conne}
\Gamma^{\prime a}_{cb}={\partial\phi^{r}\over\partial\phi^{\prime c}}(
{\partial\phi^{\prime a}\over\partial\phi^{l}}\Gamma^{l}_{rm}
{\partial\phi^{m}\over\partial\phi^{\prime b}}+{\partial^{2}\phi^{i}\over
\partial\phi^{r}\partial\phi^{\prime b}}{\partial \phi^{\prime a}\over\partial
\phi^{i}})
\ee
Usually people associate a connection with a space endowed with a metric
but this is not a necessary condition. In fact connection can also be
built\scite{fedo} out of a space equipped only with  a symplectic structure 
$\omega$ as our space ${\cal M}$ is. The difference with metric spaces is that,
while there the requirement that the connection preserve the metric scalar 
product between vectors plus that it is torsionless
makes the connection {\it unique}, in symplectic spaces the 
analogue requirement
that the connection preserves the symplectic scalar product\footnote{By symplectic
scalar product between two vectors $V^{a}$ and $W^{b}$ we mean: $V
W\equiv V^{a}\omega_{ab}W^{b}$.} does not make the
connection unique\scite{fedo}. Even if not unique what is important
is that "{\it there is a symplectic connection on any symplectic manifold}"
(see Proposition (2.5.2) on pag.66 of ref.\scite{fedo}). Few more details
are contained in the appendix.
\par
Using a symplectic connection we can now build the following object which is
a generalization of the variable $\lambda$:

\be
\label{eq:lam}
\Lambda_{a}\equiv\lambda_{a}+ic^{l}\Gamma^{m}_{al}{\bar c}_{m}
\ee
and we can prove (see the calculations in the appendix ) 
that $\Lambda$, differently than $\lambda$ (see eq.(\ref{eq:uffa})), 
transforms as a vector under a 
diffeomorphism generated by a vector field $\varepsilon^{a}(\phi)$, i.e:
\be
\label{eq:lamcor}
\Lambda_{a}^{\prime}=\Lambda_{a}+\Lambda_{b}\partial_{a}{\epsilon}^{b}
\ee
\par
One more  thing that we would like to study
is the manner to rewrite the Lie-derivative ${\widetilde{\cal H}}_{V}$ 
(\ref{eq:lie-deriv}) using the variables $\Lambda_{a}$. We will restrict
ourselves to Lie-derivatives of an Hamiltonian flow where $V^{a}\equiv
\omega^{ab}\partial_{b}H$ with $H$ a function in phase-space but not
necessarily the Hamiltonian of the system. The reason we want to rewrite
${\widetilde{\cal H}}_{V}$ using $\Lambda$ is because
${\widetilde{\cal H}}_{V}$ was a scalar despite the fact that it was built
out of objects like $\lambda_{a}$ which did not have a tensorial character.
It seems more natural to write ${\widetilde{\cal H}}_{V}$ out of objects 
which have a clear tensorial character as $\Lambda$.
Let us first rewrite ${\widetilde{\cal H}}_{V}$ of eq.(\ref{eq:lie-deriv})
in the following way:
\be
\label{eq:girata}
{\widetilde{\cal H}}_{V}\equiv\lambda_{a}V^{a}-ic^{b}
\partial_{b}V^{a}{\bar c}_{a}
\ee
Here we have just exchanged $c$ with ${\bar c}$ with respect to
eq.(\ref{eq:lie-deriv}). The difference between this
${\widetilde{\cal H}}_{V}$ and the one of eq.(\ref{eq:lie-deriv})
is zero because the extra piece which would be produced is
$\partial_{a}V^{a}$ which, with $V^{a}=\omega^{ab}\partial_{b}H$,
is zero. This extra piece would be produced
if we think of the $c$ and ${\bar c}$ as acting inside the path-integral
and so having an anticommutator different from zero.
Let us now replace the $\lambda_{a}$ in (\ref{eq:girata}) with the $\Lambda_{a}$
using the relation (\ref{eq:lam}). We  obtain:
\bea
\label{eq:girata1}
{\widetilde{\cal H}}_{V} & = & V^{a}[\Lambda_{a}-ic^{l}\Gamma_{al}^{m}{\bar
c}_{m}]-ic^{b}\partial_{b}V^{a}{\bar c}_{a} \nonumber \\
\label{eq:girata2}
& = & V^{a}\Lambda_{a}-ic^{b}[\partial_{b}V^{a}+\Gamma_{mb}^{a}V^{m}]{\bar
c}_{a}\nonumber \\
\label{eq:girata3}
& \equiv  & V^{a}\Lambda_{a}-ic^{b}\partial_{;b}V^{a}{\bar c}_{a}
\eea

In the last line of the equation above we have introduced the covariant
derivative  on vectors $V^{a}$ which is defined \scite{novi} as 
$\partial_{;b}V^{a}\equiv \partial_{b}V^{a}+\Gamma_{mb}^{a}V^{m}$.
The covariant derivative  has the well-known\scite{novi} property that 
$\partial_{;b}V^{a}$ 
transform as a tensor under diffeomorphism of $\phi$ while the usual
derivative $\partial_{b}V^{a}$ does not.
Looking at eq.(\ref{eq:girata3}) we see that ${\widetilde{\cal H}}_{V}$
has the same form as the ${\widetilde{\cal H}}_{V}$ of eq.(\ref{eq:girata})
but with the replacements:
\bea
\label{eq:sosti1}
\lambda_{a} & \Longrightarrow & \Lambda_{a}\\
\label{eq:sosti2}
\partial_{a} & \Longrightarrow & \partial_{;a}
\eea
These substitutions replace non-covariant expressions with covariant ones. As
the ${\widetilde{\cal H}}_{V}$ was a covariant quantity (a scalar), it was
natural to expect that it would not change at all under this replacement
and in fact the ${\widetilde{\cal H}}_{V}$ in eq.(\ref{eq:girata3})
is the same, even if written with covariant quantities, as the one in
(\ref{eq:girata}).
\par
Using this analysis we could ask the question of when is that an observable
${\cal O}(\phi,\lambda, c, {\bar c})$ is a scalar. The answer is the following:
take ${\cal O}$ and use the relation (\ref{eq:lam}) to replace $\lambda$
with $\Lambda$. If what you get has the same functional form in $\Lambda$
as ${\cal O}$ had in $\lambda$, with at most the ordinary derivatives replaced
by covariant derivatives, then ${\cal O}$ is a scalar.
\par
Let us now turn to the initial issue of building symmetric tensors. This is something we failed
in doing \scite{En2} by using only $\lambda$. The only thing we succeeded
was in building\scite{En1} anti-symmetric tensors using the ${\bar c}$.
Let us note   that, besides transforming correctly as  vectors,
the $\Lambda_{a}$ also commute  among themselves.
Thanks to these properties we have a manner to build symmetric tensors as
 
\be
\label{eq:TENS}
{\cal T}\equiv {\cal T}^{ab\cdots n}{\Lambda_{a}}{\Lambda_{b}}\cdots
{\Lambda_{n}}
\ee 
This\footnote{ The reader should remember that here we mean "commuting" 
among the $\Lambda$ not in an operatorial sense  but in the sense of 
considering $c$ and ${\bar c}$ entering the $\Lambda$ as
Grassmannian variables which anticommutes among themselves. The
${\bar c}$ acquires the operatorial meaning of being the ${\partial\over
\partial c}$ only once it is inserted into the path-integral. Here,instead,
in building the tensors ${\cal T}$ of eq.(\ref{eq:TENS}) we do not need to use
the path-integral at all, that is why the $\Lambda$ "commutes" among themselves.
} settles one 
of the points which was not clear in \scite{En1}.
What seems really impossible in our formalism is a manner to build
{\it mixed} tensors. The reader may be tempted to build strings
of objects containing both ~$\Lambda$~ and ${\bar c}$ like
${\cal P}={\cal P}^{abc\cdots lmn}\Lambda_{a}\Lambda_{b}\Lambda_{c}\cdots
{\bar c}_{l}{\bar
c}_{m}{\bar c}_{n}$, but this is anti-symmetric in the exchange of the 
${\bar c}$
among themselves and symmetric in the exchange the $\Lambda$ among themselves
and with the ${\bar c}$ so it is not a mixed tensor.

\section{Geometric Structure.}
In this section we will try to understand which kind of space is the one 
labeled by our 8n coordinates $(\phi,\lambda,c,\bar c)$ and correct
some wrong statements present in \scite{En1}\scite {En2}.
\par
As we said at the end of section 2 the {\it base-space} to be considered is the one labeled by
$(\phi, c)$. Let us then first
find out which kind of space is this. $\phi^{a}$ are the 2n coordinates
of the phase-space ${\cal M}$. The $c^{a}$ transform under a
diffeomorphism (see eq.(16)) as the forms $d\phi^{a}$. So we stated
in refs.\scite{En1}\scite{En2}
that, {\it identifying} $c$ with $d\phi$, the space
$(\phi^{a}, c^{a})$ makes up the cotangent bundle\scite{Mrs}
to phase space: $T^{\star}{\cal M}$. That is {\it wrong} !. In fact
$c^{a}$ are at most {\it a basis} in the fiber $T_{\phi}^{\star}{\cal M}$
and not a generic vector in $T_{\phi}^{\star}{\cal M}$.
Being the $c$ a basis they belong to what is called\scite{koba}
the {\it bundle of linear frames} to\footnote{Actually the 
bundle-of-linear-frames is made out of basis of the tangent fibers while
ours is of cotangent fibers, but the two are isomorphic.}
{\cal M}. So the $(\phi^{a},c^{a})$ are nothing else than a {\it section}
of the linear-frame-bundle\footnote{That way to look at our variables
is the one we basically adopted, without realizing it, in refs.\scite{En1}\scite{En2} and in the
previous three sections of this paper.}. We say a section because there are other
basis (or frames), besides $c^{a}$, which one could choose in the fibers of the
linear-frame-bundle.
\par
As we stressed the structure above holds if one {\it identifies} $c^{a}$ with the $d\phi^{a}$.
We did that identification in ref.\scite{En1} and used it to turn the whole
Cartan calculus into operations which could be done via our path-integral
and the structures (commutators, BRS charges, etc) present in it. Of course
the fact that the $c^{a}$ transforms as the $d\phi^{a}$ does not force
us to identify them so explicitly as we have done in ref.\scite{En1}. For
example if we build a generic vector field 
$V\equiv V^{a}{\partial\over\partial\phi^{a}}$,
we would have that the components $V^{a}$ transforms as the $c^{a}$, so we
could do this identification and the $c^{a}$ would then be {\it components}
(and not basis)  of the vectors in the tangent fibers and they 
would make up with the
$\phi$ the tangent bundle to phase-space $T{\cal M}$.
The problem is that the $c^{a}$ have a Grassmannian nature and not like
the $V^{a}$ a bosonic one. In that case the bundle is 
called\scite{Scw} {\it reversed parity} tangent bundle and indicated as 
$\Pi T{\cal M}$.
\par
Next we have to consider the role of the 4n remaining
variables $(\lambda_{a},{\bar c}_{a})$. Looking at the Lagrangian
in (\ref{eq:lagran}) we see that they play the role of momenta to
the variables $(\phi^{a},c^{a})$, so they will make the cotangent fibers
to the previous space. We can summarize all this in the following scheme:
\bea
\label{eq:schema}
(\phi^{a}) & \Longrightarrow & \cal M \\
\label{eq:schema1}
(\phi^{a},c^{a}) & \Longrightarrow & \Pi T{\cal M} \\
\label{eq:schema2}
(\phi^{a},c^{a},\lambda_{a},{\bar c}_{a}) & \Longrightarrow & T^{\star}
(\Pi T{\cal M})
\eea
On the other hand in ref.\scite{En1} (eq.3.33) we proved 
that the~${\bar c}_{a}$~act, in the operatorial counterpart of our 
path-integral, as ${\bar c}_{a}={\partial\over\partial c^{a}}$. Moreover
in the previous section of this paper we proved that
$\lambda_{a}$ , despite their strange transformation properties
(\ref{eq:lambda-giusto}), still maintain their operatorial meaning of being:
$\lambda_{a}=-i{\partial\over\partial \phi^{a}}$. From these two expressions
we can say that $(\lambda_{a},{\bar c}_{a})$ form {\it a basis} in the 
tangent fibers to the {\it base-space} $(\phi^{a},c^{a})$. As this base-space
was $\Pi T{\cal M}$, the over-all 8n coordinates $(\phi^{a},c^{a},
\lambda_{a},{\bar c}_{a})$   can also be considered as a section
of the bundle of linear frames over $\Pi T{\cal M}$. This is an 
{\it alternative}
interpretation of our 8n variables with respect to the interpretation
contained in  eq.
(\ref{eq:schema2}). This sort of "duality" between considering each of our
variables $(\lambda, c,{\bar c})$ either as {\it basis} or as {\it coordinate}
could be considered at each of the levels of
eqs.(\ref{eq:schema1})(\ref{eq:schema2}) and gives
rise to all possible combinations. We will anyhow stick  here to the {\it
"coordinate"}-picture 
 which will lead to the "reversed-parity" bundle of
eq.(\ref{eq:schema2}).
\par
Beside this sort of "duality" which would allow us to see in two different
ways the spaces labeled by our variables, there is a further freedom.
This is related to the scheme of eqs.(\ref{eq:schema1})(\ref{eq:schema2}).
Let us perform a partial integration in the kinetic piece
of the Grassmannian variables present in the Lagrangian (\ref{eq:lagran}).
Modulo surface terms the new Lagrangian is
\be
\label{eq:lagran1}
{\cal L}^{\prime}\equiv\lambda_{a}[{\dot\phi}^{a}-\omega^{ab}
\partial_{b}H]-i{\dot {\bar c}}_{a}c^{a}-i{\bar c}_{a}\omega^{ac}\partial_{c}
\partial_{b}Hc^{b}
\ee
Being this Lagrangian different from ${\widetilde{\cal L}}$ only by a surface term, 
the equations of motion for $c^{a}$ and ${\bar c}_{a}$ are the same,
but now ${\bar c}_{a}$ plays the role of a "configurational" variable
while $c^{a}$ is its relative momentum. Then it would seem natural to
choose in eq.(\ref{eq:schema1}) as new variables $(\phi^{a},{\bar c}_{a})$.
Let us see its geometrical interpretation: the ${\bar c}_{a}$ transform
as (\ref{eq:trans2}) so, interpreting the ${\bar c}_{a}$ as "coordinates" 
and not "basis", they transform as components of  forms but with Grassmannian
character, i.e. with the reversed parity character. This means  that
the $(\phi^{a},{\bar c}_{a})$ are the coordinates of the reversed parity
cotangent bundle:
\be
\label{eq:regiu1}
(\phi^{a},{\bar c}_{a})\Longrightarrow \Pi T^{\star}M
\ee
From the Lagrangian ${\cal L}^{\prime}$ of (\ref{eq:lagran1}) we see that
$\lambda_{a}$ and $c^{a}$ play the role of momenta to the previus
variables so they belong to the cotangent fibers of the previous space.
All together then we can write all this in the following scheme
\bea
\label{eq:schemo}
(\phi^{a}) & \Longrightarrow & \cal M \\
\label{eq:schemo1}
(\phi^{a},{\bar c}_{a}) & \Longrightarrow & \Pi T^{\star}{\cal M} \\
\label{eq:schemo2}
(\phi^{a},{\bar c}_{a},\lambda_{a},c^{a}) & \Longrightarrow & T^{\star}
(\Pi T^{\star}{\cal M})
\eea
As the physics contained in the Lagrangians ${\cal L}$ and ${\cal L}^{\prime}$
is the same and the same are the coordinates, we could say that our
variables label both spaces either $T^{\star}(\Pi T{\cal M})$ or 
$T^{\star}(\Pi T^{\star}{\cal M})$. A more mathematically precise
proof of this is contained in ref.\scite{mregini}.

\section{Entirely Bosonic Path Integral.}

The reader may feel a little been uneasy with these Grassmannian double
bundle we have provided in the previous section
and  even with the alternative interpretation as sections of 
the frame bundle. For this reason in this section we will show that, 
at least for {\it Hamiltonian} flow, it is possible to provide a path-integral 
of classical mechanics
made entirely of bosonic variables. Moreover we will prove that these
variables are just  the coordinates of a standard double
bundle like  $T^{\star}(T^{\star}{\cal M})$.
The procedure to achieve what we said above is explained below.
\par
The path-integral\scite{En1} for {\it classical} mechanics (CM) was basically
the following. We wanted to create a generating functional $Z[J]$  
which would give
to each path not the Feynman weight $exp {i\over\hbar}S$ but weight 1 if 
the path
was a classical one and weight zero if it was a non-classical one. So essentially:
\be
\label{eq:bos1}
Z[J]=\int{\cal D}\phi~{\tilde\delta}[\phi(t)-\phi_{cl}(t)]~exp\int J\phi dt
\ee
where $\phi_{cl}$ are the solutions of the Hamiltonian equations of motion:
$\dot\phi^{a}=\omega^{ab}{\partial H\over\partial\phi^{b}}$ and the
${\tilde\delta}[\cdot]$ is a functional Dirac delta.
It is easy to realize\scite{En1}, neglecting the current $J$  for a moment,
we can rewrite (\ref{eq:bos1}) as
\bea
\label{eq:bos2}
Z[J]= \int{\cal D}\phi~{\tilde\delta}[\phi(t)-\phi_{cl}(t)] & = & \int {\cal D}
\phi~{\tilde\delta}[{\dot\phi}^{a}-\omega^{ab}{\partial H\over\partial\phi^{b}}]
~det[\delta^{a}_{l}\partial_{t}-\omega^{ab}{\partial^{2}H\over\partial\phi^{b}
\partial\phi^{l}}]  \\
\label {eq:bos3} 
& = & \int {\cal D}\phi^{a}{\cal D}\lambda_{a}{\cal D}c^{a}{\cal D}{\bar c}_{a}
exp~i\int {\widetilde{\cal L}}dt
\eea
where the $det[\cdot]$ appearing in eq.(\ref{eq:bos2}) is a functional
determinant\scite{En1} and the ${\widetilde{\cal L}}$ in
(\ref{eq:bos3}) is 
the Lagrangian of eq.(\ref{eq:lagran}). It is  obtained by doing the Fourier transform
(via the variables $\lambda_{a}$) of the Dirac delta in the second term of 
eq.(\ref{eq:bos2}) and exponentiating the $det[\cdot]$ with Grassmannian
variables $c^{a}$ and ${\bar c}_{a}$.
In order to avoid using the Grassmannian variables the trick we adopt now
is to substitute the $det[\cdot]$ in (\ref{eq:bos2}) with an inverse 
determinant:
\be
\label{eq:inve}
det[\delta^{a}_{l}\partial_{t}-\omega^{ab}{\partial^{2} H\over\partial\phi^{b}
\partial\phi^{l}}]=\{det[\delta^{a}_{l}\partial_{t}+\omega^{ab}{\partial^{2} H
\over\partial\phi^{b}\partial\phi^{l}}]\}^{-1}
\ee
We will give a detailed proof of this relation in the appendix.
\par
The next step is to use relation (\ref{eq:inve}) into (\ref{eq:bos2}),
then "exponentiate" the inverse of the matrix using bosonic variables 
by making use of the well known formula\footnote{This formulas 
requires that the determinant be positive and this is our case
because the LHS of (\ref{eq:inve}) is positive\scite{En1}.}:

\be
\label{eq:gauss}
\int dx^{i}dy_{j}~exp~ix^{i}A_{i}^{j}y_{j}\propto \{det[A_{i}^{j}]\}^{-1}
\ee
Doing all that we get 
\bea
\label{eq:bos4}
Z[J] & = & \int {\cal D}
\phi {\tilde\delta}[{\dot\phi}^{a}-\omega^{ab}{\partial H\over\partial\phi^{b}}]
\{det[\delta^{a}_{l}\partial_{t}+\omega^{ab}{\partial^{2}H\over\partial\phi^{b}
\partial\phi^{l}}]\}^{-1}  \\
\label {eq:bosi3} 
& = & \int {\cal D}\phi^{a}{\cal D}\lambda_{a}{\cal D}\pi^{a}{\cal D}{\xi}_{a}
exp~i\int {\cal L}dt
\eea
where 
\be
\label{eq:bosu}
{\cal L}=\lambda_{a}[{\dot\phi}^{a}-\omega^{ab}{\partial
H\over\partial\phi^{b}}]+\pi^{l}[\delta^{a}_{l}\partial_{t}+\omega^{ab}
{\partial^{2}H\over\partial\phi^{b}\partial\phi^{l}}]\xi_{a}
\ee
The variables $\pi^{l}$ and $\xi_{a}$ are the Bosonic variables we have
used to exponentiate the inverse matrix and they replace the Grassmannian
variables $c^{a}$ and ${\bar c}_{a}$ present in ${\widetilde{\cal L}}$
of eq.(\ref{eq:lagran}).
\par
Let us now see if we can give a geometrical understanding of the new
variables $\pi^{a}$,$\xi_{a}$ present here. Let us show how they change
under the Hamiltonian flow, that means under their equation of motion 
which can easily be derived from the Lagrangian ${\cal L}$ above:
\be
\label{eq:bosu2}
\partial_{t}\xi_{l}+\xi_{a}\omega^{ab}{\partial^{2}H\over\partial\phi^{b}
\partial^{l}}  =  0
\ee
This  equation should be compared with the equations of motion
of $c^{a}$ derived\scite{En1} from ${\widetilde{\cal L}}$ of
eq.(\ref{eq:lagran}) which are
\be
\label{eq:bosu3}
\partial_{t}c^{a}-\omega^{ab}{\partial^{2}H\over\partial\phi^{b}
\partial\phi^{l}}c^{l}  =  0
\ee

From the above equations it is now easy to see that the quantity
\be
\label{eq:marc}
\Xi\equiv \xi_{a}c^{a}
\ee is
invariant under the Hamiltonian flow.
This quantity would behave in the same way
under any diffeomorphism
of the phase-space~${\cal M}$ and not just under the Hamiltonian
flow\footnote{This is so because we would have to choose
the transformations on $\pi$ and $\xi$ induced by the diffeomorphism
in $\phi$ in such a way to keep invariant
the Hamiltonian associated to ${\cal L}$.}. 
The invariance of $\Xi$ is the same thing that would happen to a form\footnote{
This is so because forms are object totally coordinate free.} 
\be
\label{eq:marc1}
{\tilde\Xi}\equiv{\tilde\xi}_{a}d\phi^{a}
\ee
and by identifying the
$d\phi^{a}$ above  with the $c^{a}$ of eq.(\ref{eq:marc})
we see that we can identify $\xi_{a}$ of (\ref{eq:marc})
with the components ${\tilde\xi}_{a}$ of the forms of (\ref{eq:marc1}).
So while the
$c^{a}$ are the basis of the fibers on $T^{\star}{\cal M}$, the
$\xi_{a}$ are the coordinates of the same space. Looking at the Lagrangian (\ref{eq:bosu})
we see that $\pi^{a}$ and $\lambda_{a}$ are the momenta associated
to $\phi^{a}$ and $\xi_{a}$, that means they will make up the cotangent
fibers to the previous space,
So the overall set of variables $(\phi^{a},\xi_{a},\lambda_{a},\pi^{a},)$
are the coordinates of $T^{\star}(T^{\star}{\cal M})$. This is a double bundle
but it may please more the reader that the reversed parity one 
$T^{\star}(\Pi T{\cal M})$ associated to the Lagrangian 
${\widetilde{\cal L}}$ of (\ref{eq:lagran}).
It may be a space easier to handle for the study of various {\it physical} issues
like the study of ergodicity and Lyapunov exponents \scite{En3} we performed
previously using the old Lagrangian (\ref{eq:lagran}). It may also be worth
to see if the universal symmetries (BRS and Supersymmetry) we found in \scite
{En1}\scite{En3} are present (in a different form) also in this purely bosonic
case presented here.
\section{Conclusions}
In this paper we have settled several geometrical issues still opened
in the path-integral approach to classical mechanics. We think that the light
shed on the "dual" aspects of the  geometrical interpretations of the 
Grassmannian variables is something important and 
long overdue. At the same time we have disentangled the puzzles related
to the $\lambda$ variables understanding the manner to  build symmetric 
tensors. Last but not least we have to say that having cleared
all the geometry involved has helped us in our search
for a purely bosonic path-integral.
Overall we think that it
was not useless to do all this work especially  considering the relevant role 
that geometry is having  in physics in the last ten years.

\newpage

\section*{Appendix }
In this appendix we provide some calculational details which will make
the paper self-contained.
\par
$\bullet$ As we said in section 3 there are various
extra conditions which one could impose on the symplectic connections.
One condition, which anyhow will not make the connection unique, 
is the requirement
that the connection be torsionless which is equivalent to saying
that it must be  symmetric in the two lower indices\scite{novi}:~
$\Gamma_{ab}^{c}=\Gamma_{ba}^{c}$. For the reader not familiar with torsion
we can phrase this requirement in the following other form:
having a connection one can build a covariant derivative on the space of forms.
It is  easy to see that, if we write forms as 
$ { \cal F }={ \cal F }_{ab\cdots m}c^{a}c^{b}\cdots c^{m}$, 
the covariant derivative  acting on them is
\be
\label{eq:cov}
\nabla_{a}\equiv{\partial\over\partial\phi^{a}}-c^{l}\Gamma^{m}_{al}{\partial
\over\partial c^{m}}
\ee
Using it we could then obtain a {\it new} "exterior derivative". The old one
was\scite{En1} $d=c^{a}{\partial\over\partial\phi^{a}}$ and the new one
would be ${\tilde d}\equiv
c^{a}\nabla_{a}=c^{a}[{\partial\over\partial\phi^{a}}-c^{l}
\Gamma^{m}_{al}{\partial
\over\partial c^{m}}]$. Here one see immediately that if $\Gamma$ were
symmetric in the two lower indeces, the ${\tilde d}$ would turn into the old
exterior derivative $d$. So requiring this symmetry in the lower
indices would not only make our theory torsionless but it would also prevent 
the appearance of two different exterior derivatives.
\par
$\bullet$
As we promised after eq.(\ref{eq:lam}) we will show here that  $\Lambda$
transforms as a vector under a diffeomorphism of the phase-space
: $\phi^{a}\rightarrow \phi^{\prime a}(\phi)$.  Let us first write
$\Lambda$ as $\Lambda_{a}=\lambda_{a}+i W_{a}$ where (see eq.(\ref{eq:lam}))
$W_{a}\equiv c^{l}\Gamma_{al}^{m}{\bar c}_{m}$. We know how $\lambda$
transforms under the diffeomorphism indicated above
 (see eq.(\ref{eq:coord-free})) and so we have only to find how $W$
changes under the same transformations. Using the transformations rules
for $\Gamma$ (eq.(\ref{eq:conne})) and for $c$ and ${\bar c}$ which are:

\bea
\label{eq:basta}
c^{\prime a} & = & {\partial\phi^{\prime a}\over\partial\phi^{b}}c^{b} \\
\label{eq:basta1}{\bar c}^{\prime}_{a} & = & {\partial\phi^{b}\over
\partial\phi^{\prime a}}{\bar c}_{b} 
\eea
It is a simple but long calculation
to show how the $W_{a}$ transform:
\bea
\label{eq:regis5}
W^{\prime}_{a} & = & {\partial\phi^{r}\over\partial\phi^{\prime a}}W_{r}+
{\partial^{2}\phi^{j}\over\partial\phi^{\prime a}\partial \phi^{\prime l}}
{\partial\phi^{\prime l}\over\partial\phi^{k}}c^{k}{\bar c}_{j}\nonumber\\
\label{eq:regiss5}
& = & {\partial\phi^{r}\over\partial\phi^{\prime a}}W_{r}+{\partial c^{s}\over
\partial \phi^{\prime a}}{\bar c}_{s}
\eea
Combining the equation above with (\ref{eq:coord-free}) we get that:
\newpage
\bea
\label{eq:regis6}
\Lambda^{\prime}_{a} & \equiv & \lambda^{\prime}_{a}+iW^{\prime}_{a}\nonumber \\
\label{eq:regiss6}
& = & {\partial\phi^{b}\over\partial\phi^{\prime a}}\lambda_{b}-i{\partial
c^{b}\over\partial\phi^{\prime a}}{\bar c}_{b}+i{\partial\phi^{b}\over
\partial\phi^{\prime a}}W_{b}+i{\partial c^{b}\over\partial\phi^{\prime a}}
{\bar c}_{b}\nonumber\\
\label{eq:regiss7}
& = & {\partial\phi^{b}\over\partial\phi^{\prime a}}\lambda_{b}+i{\partial
\phi^{b}\over\partial\phi^{\prime a}}W_{b}=
{\partial\phi^{b}\over\partial\phi^{\prime a}}\Lambda_{b}
\eea
If the transformation is generated by a vector field $\varepsilon^{a}(\phi)$,~
i.e.:$\phi^{\prime a}=\phi^{a}-\varepsilon^{a}(\phi)$, then eq.(\ref{eq:regiss7})
becomes exactly eq.(\ref{eq:lamcor}).
Eq.(\ref{eq:regiss7}) proves that $\Lambda$ transforms as a vector. Somehow the extra
pieces in the transformations of $\lambda$ (eq.(\ref{eq:coord-free}))
got cancelled by the extra pieces in the transformations of $W$ (in
eq.(\ref{eq:regiss5})).
\par 
$\bullet$
We will now give a {\it formal} proof of formula (\ref{eq:inve}).
The determinant in (\ref{eq:inve}) are functional determinant
that means:
\bea
\label{eq:func}
det[\delta^{a}_{l}\partial_{t}-\omega^{ab}{\partial^{2} H\over\partial\phi^{b}
\partial\phi^{l}}]& \equiv & det[\delta^{a}_{l}\partial_{t}\delta(t-t^{\prime})-
\delta(t-t^{\prime})\omega^{ab}{\partial^{2}H\over\partial\phi^{b}
\partial\phi^{l}}]\\
\label{eq:func1}
& = & \{ det \partial_{t}\}\{det[\delta^{a}_{l}\delta(t-t^{\prime})-
\theta(t-t^{\prime})\omega^{ab}{\partial^{2}H\over\partial\phi^{b}
\partial\phi^{l}}]\}
\eea
To prove (\ref{eq:inve}) is equivalent to saying that the determinant
of the product of the two matrices appearing respectively on the LHS 
and RHS of (\ref{eq:inve}) is one. Using the form of the matrix written
in eq.({\ref{eq:func1}), this means\footnote{We can use that form of the matrix
because they differ by a constant factor independent of the $\phi$.}:
\bea
\label{eq:inver}
& det & \{ \int dt^{\prime}[\delta^{a}_{b}\delta(t-t^{\prime})-\theta
(t-t^{\prime})\omega^{al}{\partial^{2}H\over\partial\phi^{l}\partial\phi^{b}}]
[\delta^{b}_{c}\delta(t^{\prime}-t^{\prime\prime})+\theta
(t^{\prime}-t^{\prime\prime})\omega^{bk}{\partial^{2}H\over\partial{\phi^{k}}
\partial\phi^{c}}]\}=\nonumber \\
\label{eq:inver1}
& det & \{  \delta^{a}_{c}\delta(t-t^{\prime\prime})-\int dt^{\prime}
\theta(t-t^{\prime})\theta(t^{\prime}-t^{\prime\prime}) \omega^{al}
{\partial^{2}H\over\partial\phi^{l}\partial\phi^{b}}
\omega^{bk}{\partial^{2}H\over\partial\phi^{k}\partial\phi^{c}}\}\\
\label{eq:inver2}
& \approx & exp~-\int dt^{\prime}\theta(t-t^{\prime})
\theta(t^{\prime}-t)\omega^{al}{\partial^{2}H\over\partial\phi^{l}
\partial\phi^{b}}\omega^{bk}{\partial^{2}H\over\partial\phi^{k}
\partial\phi^{a}}=1
\eea
In (\ref{eq:inver2}) we have used the "exp-tr" form for the determinant
and  fact that the two $\theta(\cdot)$
give zero.
So this proves the relation (\ref{eq:inve}).

$\bullet$
Here we will  qualify the steps done in eqs.(74) and (75).
There we used the $\theta(t-t^{\prime})$ as "inverse" (or Green function) 
of $\partial_{t}$. The $\theta(t-t^{\prime})$  is the retarted or {\it causal}
Green function. If we had used other Green functions like the
$\epsilon(t-t^{\prime})$, we would not have obtained equation (75). 

It is actually well-known~\cite{NAKA}\cite{OKA} 
that all functional determinants of the form:
\begin{equation}
\label{eq:unorefe}
det[\partial_{t}\delta(t-t^{\prime})-\delta(t-t^{\prime})G^{\prime}(\varphi)]
\end{equation}
depend on the {\it boundary conditions} under which we solve the associated
differential equation:
\begin{equation}
\label{eq:trerefe}
[\partial_{t}-G^{\prime}(\varphi)]c_{n}(t)=\sigma_{n}c_{n}(t)
\end{equation}
whose eigenvalues $\sigma_{n}$ are needed to calculate the determinant in some
{\it regularized} form:
\begin{equation}
\label{eq:duerefe}
det[(\cdot)]=\{\prod_{n=-\infty}^{n=\infty} 
\sigma_{n}\}_{\scriptscriptstyle{regul.}}
\end{equation}
\noindent
Solving equation (77) with {\it causal} boundary conditions
 one obtains~\cite{OKA}:

\begin{equation}
\label{eq:undicirefe}
det[\partial_{t}\delta(t-t^{\prime})-\delta(t-t^{\prime})
G^{\prime}(\varphi)]_{\scriptscriptstyle causal.}=exp -{1\over 2}\int
dt^{\prime}G^{\prime}(\varphi(t^{\prime}))
\end{equation}
\noindent
So by reversing the sign of $G^{\prime}(\varphi)$ we get:

\begin{equation}
\label{eq:dodicicirefe}
det[\partial_{t}\delta(t-t^{\prime})+\delta(t-t^{\prime})
G^{\prime}(\varphi)]_{\scriptscriptstyle causal.}=exp +{1\over 2}\int
dt^{\prime}G^{\prime}(\varphi(t^{\prime}))
\end{equation}
\noindent
By comparing the RHS of the last two equations above we see that the
two determinants are one the inverse of the other. This proves relation (58)
provided we specify that the functional determinant has to be evaluated
with {\it causal} boundary conditions.

The reason we choose these boundary conditions is because, after all, we are 
just doing classical mechanics that means just solving ordinary Hamiltonian 
equations of motion. These are usually solved by giving a value of $\varphi$ at
the intial
time $t=0$ and looking for the evolution at {\it later} times using a {\it
causal} propagator.  The use of  {\it periodic b.c.} and of a time-symmetric
Green function 
for our path-integral has been analyzed in full details in ref.\cite{topo}.
The result is a path-integral whose only {\it non-zero} expectation values
are those associated to observables which are independent from deformations
of the Hamiltonian $H$ and of its symplectic form $\omega_{ab}$. This means
a path-integral which does not feel anymore the form of H, so something that
does not feel the dynamics at all and this is  is not what we want here.

%
\section*{Acknowledgments}
We are expecially grateful to D.Mauro for many discussions
and much needed help during the last stages of this work.
We are also grateful to  G.Landi, G.Marmo  for
helpful discussions and to F.Benatti who, since December 1994,
insisted that we find a purely bosonic representation of
our path-integral. This work has been supported by  grants from 
MURST (Italy)  and NATO.




\begin{thebibliography}{99}
\bi{En1}
E.Gozzi, M.Reuter and W.D.Thacker, Phys.Rev.D {\bf 40}, 3363 (1989).

\bi{En2}
E.Gozzi, M.Reuter and W.D.Thacker, Phys. Rev.D {\bf 46}, 757 (1992).

\bi{Scw}
A.Schwarz, in "{\it Topics in statistical and theoretical physics}"
Ed. R.L.Dobrushin. 

\bi{Mrs}
R.Abraham and J.Marsden, "{\it Foundations of Mechanics}" (Benjamin, New York,
1978). 

\bi{Vienna}
I.Kolar, P.W.Michor, J.Slovak, "{\it Natural Operations in Differential
Geometry}" (Springer Verlag 1993, Berlin).

\bi{novi}
S.P.Novikov, B.A.Dubrovin, A.T.Fomenko, "{\it Modern geometric methods
and applications.}" Springer-Verlag, New York, 1992.

\bi{fedo}
B.Fedosov, "{\it Deformation quantization and index theory}" Akademie Verlag,
Berlin 1996.

\bi{koba}
S.Kobayashi and K.Nomizu, "{\it Foundations of differential geometry}" Vol.I
John Wiley and Sons Editors, New York 1963

\bi{En3}
E.Gozzi and M.Reuter, Phys. Lett. B 233 (1989) 383; Chaos, Solitons and
Fractals, Vol.4, no.7 (1994) 1117.

\bi{mregini}
Marco Regini, Tesi di Laurea, Universita' di Trieste, July 1998.

\bibitem{NAKA}
H.Nakazato et al, Nucl.Phys. B346 (1990) 611-631;

\bibitem{OKA}
S.Marculescu et al., Nucl.Phys. B 349 (1991) 463-493;

\bibitem{topo}
E.Gozzi, M.Reuter, Phys. Lett. B 240 (1990) 137.

\end{thebibliography}
\end{document}